\documentclass[pra,superscriptaddress,aps,10pt,twocolumn]{revtex4-1}

\usepackage[utf8]{inputenc}
\usepackage[T1]{fontenc}
\usepackage{textcomp}
\usepackage{gensymb}
\usepackage[english]{babel}

\usepackage{amssymb,amstext}
\usepackage{amsthm}
\usepackage{amsmath}
\usepackage{amsfonts}
\usepackage{mathtools}
\usepackage{newunicodechar}

\usepackage{graphicx}
\usepackage[space]{grffile} 

\usepackage{hyperref}
\usepackage[all]{hypcap}  
\makeatletter
\hypersetup{colorlinks=true, allcolors=black}
\makeatother

\usepackage{braket}
\usepackage{url}
\usepackage{float}
\usepackage{color}

\renewcommand{\v}[1]{\vec{\boldsymbol{#1}}}
\newcommand{\vu}[1]{\hat{\boldsymbol{#1}}}
\newcommand{\dyad}[1]{|#1\rangle\langle#1|}

\begin{document}
\title{Spin-Interaction Effects for Ultralong-range Rydberg Molecules in a Magnetic Field}
\date{\today}
\author{Frederic Hummel}
\email{frederic.hummel@physnet.uni-hamburg.de}
\affiliation{Zentrum für Optische Quantentechnologien, Universität Hamburg, Luruper Chaussee 149, 22761 Hamburg, Germany}
\author{Christian Fey}
\affiliation{Zentrum für Optische Quantentechnologien, Universität Hamburg, Luruper Chaussee 149, 22761 Hamburg, Germany}
\author{Peter Schmelcher}
\affiliation{Zentrum für Optische Quantentechnologien, Universität Hamburg, Luruper Chaussee 149, 22761 Hamburg, Germany}
\affiliation{The Hamburg Centre for Ultrafast Imaging, Universität Hamburg, Luruper Chaussee 149, 22761 Hamburg, Germany}

\begin{abstract}
We investigate the fine and spin structure of ultralong-range Rydberg molecules exposed to a homogeneous magnetic field. Each molecule consists of a $^{87}$Rb Rydberg atom whose outer electron interacts via spin-dependent $s$- and $p$-wave scattering with a polarizable $^{87}$Rb ground state atom. Our model includes also the hyperfine structure of the ground state atom as well as spin-orbit couplings of the Rydberg and ground state atom. We focus on $d$-Rydberg states and principal quantum numbers $n$ in the vicinity of 40. The electronic structure and vibrational states are determined in the framework of the Born-Oppenheimer approximation for varying field strengths ranging from a few up to hundred Gauß. The results show that the interplay between the scattering interactions and the spin couplings gives rise to a large variety of molecular states in different spin configurations as well as in different spatial arrangements that can be tuned by the magnetic field. This includes relatively regularly shaped energy surfaces in a regime where the Zeeman splitting is large compared to the scattering interaction but small compared to the Rydberg fine structure, as well as more complex structures for both, weaker and stronger fields. We quantify the impact of spin couplings by comparing the extended theory to a spin-independent model.
\end{abstract}

\maketitle

\section{Introduction}

Ultralong-range Rydberg molecules (ULRM) comprising a Rydberg and a ground state atom possess huge bond lengths and can exhibit correspondingly large permanent electric dipole moments. In cold or ultracold atomic clouds, molecular bound states can be formed with an underlying oscillatory potential landscape that is induced by scattering of the Rydberg electron off the neutral ground state atom. Pictorially speaking, the potential energy surfaces (PES) due to which molecules form reflect the spatial variation of the density of the electronic Rydberg wave function. ULRM were theoretically predicted in 2000 \cite{greene_creation_2000} and come in different classes such as nonpolar molecules with low angular momenta $l\leq3$ and polar molecules for high angular momenta $l>3$, which are called trilobite. Both of these species have been found in corresponding spectroscopic experiments \cite{Bendkowsky2009, li_homonuclear_2011, Booth2015}. The electronic scattering process is modeled by a Fermi-type pseudo-potential for s-wave scattering \cite{Fermi1934}. The generalization to include p-wave scattering \cite{omont_theory_1977}, leads to a novel type of molecules, named butterfly states \cite{hamilton_shape-resonance-induced_2002}, whose existence has also been confirmed experimentally \cite{Bendkowsky2010, niederprum_observation_2016}. Both, trilobites and butterflies, share the property of a strong localization of the electronic wave function around the ground state atom, possible through a mixing of the quasi-degenerate high angular momentum $l>3$-states and allowing for a strong binding on the order of GHz. Opposite to this, low angular momentum states $l\leq3$ possess binding energies on the order of several MHz. They are energetically detuned from the hydrogenic manifold due to their non-integer quantum defect, an effect that originates from the interaction of the Rydberg electron with the core electrons of the Rydberg atom. Thus, both, localization and dipole moments, are much weaker compared to polar molecules. \\ 
For large principal quantum numbers $n\gtrsim45$, and depending on the density of the cold atomic gas, polyatomic molecules become possible, when a single Rydberg electron binds several ground state atoms \cite{liu_polyatomic_2006, liu_ultra-long-range_2009, gaj_molecular_2014, Fey2016, Eiles2016}. ULRMs have proven a valuable tool in a variety of settings from probing electron scattering resonances \cite{Schlagmuller2016} to occupancies of atomic lattices \cite{Manthey2015} and observation of ultracold chemical reactions \cite{Niederprum2015, Schlagmuller2016x}. \\
Looking at ULRM in some more detail, i.e.~for a higher resolution, further relevant interactions include spin-dependent s-wave and p-wave scattering in combination with the hyperfine structure of the ground state atom \cite{Anderson2014a}. These were very recently observed experimentally in rubidium \cite{Anderson2014, Boettcher2016} and cesium ULRMs \cite{Sassmannshausen2015}. An actual spin transfer process from the Rydberg atom to the ground state atom has been observed in reference \cite{niederprum_rydberg_2016}. For strontium with its vanishing nuclear spin, the corresponding couplings are less prominent, but even these species have been detected \cite{Camargo2016, Whalen2017}. \\
The effects of spin orbit coupling in the electron-neutral interaction of the Rydberg electron and the ground state atom were treated in an alternative theoretical approach using Green's function techniques \cite{Khuskivadze2002}. However, in this treatment the inclusion of the hyperfine structure remains unclear. Only very recently efforts have been made to incorporate the latter into the pseudo-potential approach \cite{Markson2016, Eiles2017}. \\
As a matter of fact, the electronic Rydberg wave function is very sensitive to external static or time-dependent fields. Thus, the density and consequently the potential energy surface can be tuned to support certain molecular equilibrium geometries using only weak magnetic or electric fields. Indeed, the impact of magnetic and electric fields has been studied both, theoretically \cite{Lesanovsky2006, kurz_electrically_2013} and experimentally \cite{krupp_alignment_2014, Gaj2015}, including the case of combined electric and magnetic fields \cite{kurz_ultralong-range_2014}. \\
In the present work we go one step further and combine a detailed theoretical description of ULRM including the effects of spin couplings with the application of external homogeneous magnetic fields. Specifically, we will focus on $l=2$ i.e.~$d$-state Rydberg molecules, including also higher partial wave interactions. Three regimes of magnetic field strength relative to other relevant interaction strengths emerge. In the intermediate regime, the PES are regularly shaped and effects of mixed singlet and triplet scattering are visible. For weaker fields, PES of different angular momenta projection states strongly mix influencing and distorting the shapes of individual potential wells as well as the overall PES. For stronger fields, PES of different total angular momenta states of the Rydberg electron mix, leading to a tunable molecular orientation relative to the magnetic field. \\
This work is organized as follows. Section \ref{Theory} provides the theoretical framework for the description of ultralong-range Rydberg molecules. In section \ref{Fieldfree} we briefly recapitulate the PES for the Hamiltonian developed in \cite{Eiles2017}, which is here modified by the inclusion of additional magnetic fields. In section \ref{ULRM} we present the results for $d$-state molecules in a homogeneous magnetic field, which are organized into an intermediate- (\ref{intermediate}), weak- (\ref{weak}), and strong-field regime (\ref{strong}). Section \ref{conc} contains our conclusions.

\section{Theoretical description of Rydberg molecules \label{Theory}}

\begin{figure}
\centering
\includegraphics[scale=0.95]{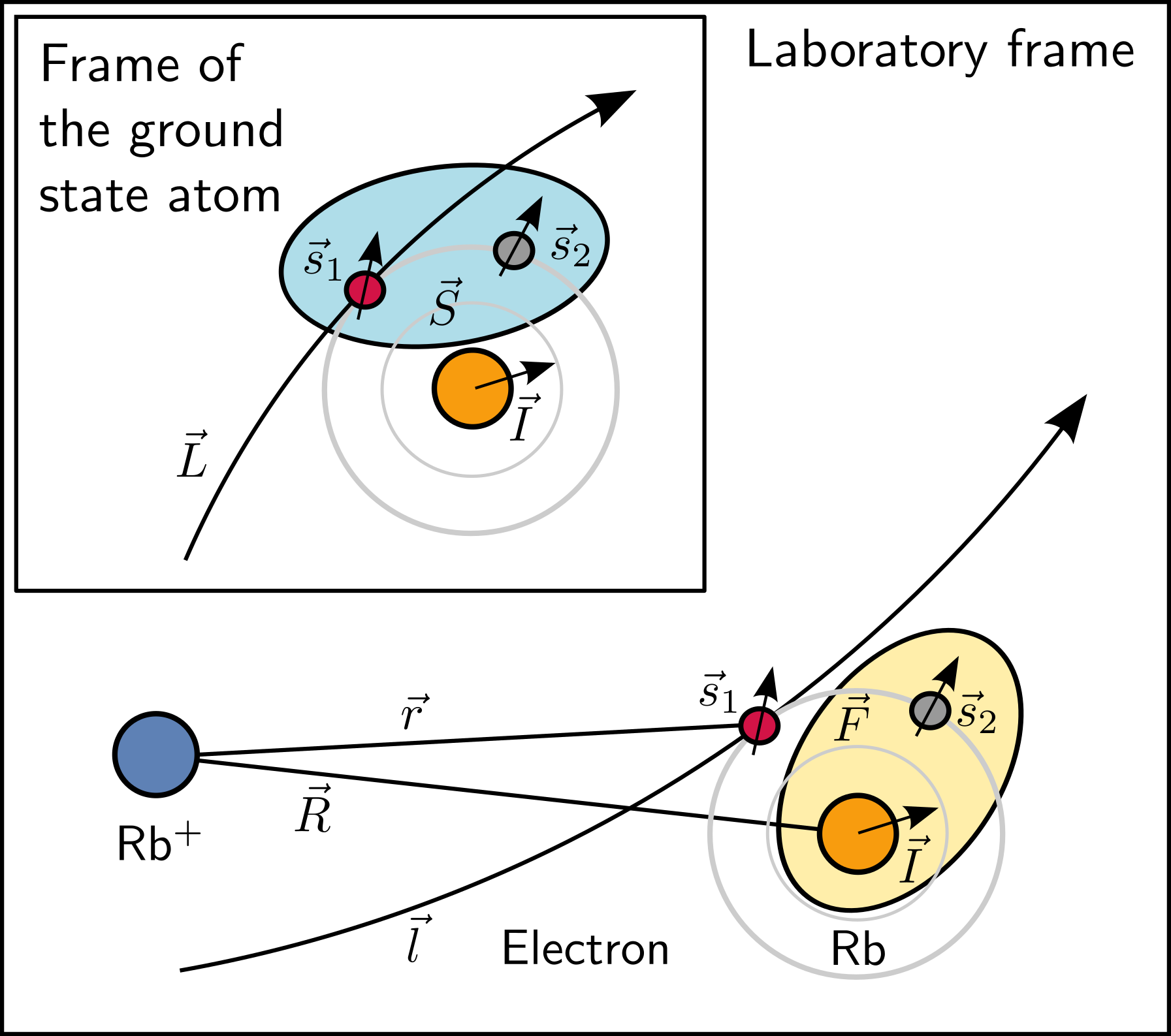}
\caption{Sketch of the molecular system. The Rydberg electron at position $\v{r}$ relative to the rubidium core carries an orbital angular momentum $\v{l}$ and spin $\v{s}_1$. The neutral ground state atom at position $\v{R}$ carries a nuclear spin $\v{I}$ which interacts with the spin $\v{s}_2$ of its valence electron to form a hyperfine state characterized by the quantum number $F$. Upon contact of the two systems (in the inset box) the total electronic spin $\v{S} = \v{s}_1 + \v{s}_2$ interacts with the angular momentum of the Rydberg electron relative to the ground state atom $\v{L}$. \label{sketch}}
\end{figure}

We consider a system of two rubidium atoms, one in a Rydberg state and the second one in its ground state at position $\v{R}$ relative to the ionic core of the Rydberg atom as depicted in figure \ref{sketch}. In the Born-Oppenheimer approximation the electronic motion is described by the time-independent Schrödinger equation
\begin{equation}
\hat{H}(\v{r};\v{R})\Psi_i(\v{r};\v{R}) = \epsilon_i(\v{R})\Psi_i(\v{r};\v{R})
\end{equation}
for the electronic wave functions $\Psi_i(\v{r};\v{R})$. $\epsilon_i(\v{R})$ are the adiabatic potential energy surfaces which depend parametrically on the relative vector $\v{R}$. In atomic units the electronic Hamiltonian is of the form \cite{Eiles2017}
\begin{equation}
\hat{H}(\v{r};\v{R}) = \hat{H}_{\text{Ryd}}(\v{r}) + \hat{H}_{\text{HF}} + \hat{V}_{\text{P}}(\v{R},\v{r}) - \frac{\alpha}{2R^4} + \hat{H}_{\text{B}}.
\end{equation}
Here, $\hat{H}_{\text{Ryd}}(\v{r})$ is the Hamiltonian of the Rydberg atom including its spin-orbit coupling. Its eigenfunctions are 
\begin{equation}
\Psi_{nljm_j}(\v{r}) = \sum_{m_l, m_1} C_{lm_l,s_1m_1}^{jm_j} \frac{1}{r} f_{nlj}(r) Y_{lm}(\hat{r}) \chi_{m_1}^{s_1},
\end{equation}
where $C_{lm_l,s_1m_1}^{jm_j}$ is a Clebsch-Gordan coefficient, $f_{nlj}(r)$ is a Whitaker function, $Y_{lm}(\hat{r})$ is a spherical harmonic, and $\chi_{m_1}^{s_1}$ is the Rydberg electron's spin wave function, with the Rydberg electron's spin $s_1$ and its projection quantum number $m_1$. Whitaker functions approximate the radial solution of the Hamiltonian by means of the asymptotic energy
\begin{equation}
E_{nlj} = -\frac{1}{2(n-\mu(n,l,j))^2},
\end{equation}
where $\mu(n,l,j)$ are the non-integer non-zero quantum defects for $l\leq 3$. The quantum defects take into account the scattering of the Rydberg electron from the Rydberg core and inner shell electrons, leading to an effective energy shift of the Rydberg state. Thus, $f_{nlj}(r)$ is a good approximation for distances larger then a few Bohr radii. For higher angular momentum the quantum defect is negligible and the Whitaker functions coincide with hydrogenic eigenfunctions. We obtain the quantum defect from experimental data of atomic spectroscopy \cite{Goy1982}. Employing the good quantum numbers $\ket{nljm_j}$ of the Rydberg electronic wave function in the absence of a perturbing ground state atom diagonalizes $\hat{H}_{\text{Ryd}}(\v{r})$. These are the principal quantum number $n$, the orbital angular momentum $l$, the total angular momentum $j$ including spin, and its projection onto the internuclear axis $m_j$. \\
The Hamiltonian of the ground state atom contains the hyperfine interaction of its nuclear and electronic spins 
\begin{equation}
\hat{H}_{\text{HF}} = A\hat{\v{I}} \cdot \hat{\v{s}}_2,
\end{equation}
with a hyperfine coupling for Rb(5s) of $A=3.417$ GHz. As a basis we choose here $\ket{s_2m_2Im_I}$, with the valence electron spin $s_2$ and the z-component of its magnetic quantum number $m_2$ as well as the nuclear spin $I$ and its projection onto the z-axis $m_I$. For rubidium the nuclear spin is given by $I=3/2$. Due to the coupling to the total nuclear spin $F=I\pm s_2$, $\hat{H}_{\text{HF}}$ is not diagonal in this basis. \\
The Fermi-like pseudo-potential $\hat{V}_{\text{P}}(\v{R},\v{r})$ models the interaction of the Rydberg electron and the neutral ground state atom \cite{Fermi1934, omont_theory_1977}, generalized to include all scattering channels up to p-wave which means spectroscopically that we account for ${}^{2S+1}L_J = {}^1S_0, {}^3S_1, {}^1P_1, {}^3P_{0,1,2}$. These channels arise because the Rydberg electron and the valence electron of the ground state atom form a singlet or triplet state $S$, which interacts via spin-orbit coupling with the electronic orbital angular momentum $L$ relative to the ground state atom to form the total angular momentum $J$ relative to the ground state atom. Each scattering channel can be associated with a scattering length or volume $a(S,L,J,k)$, which is modeled using the phase shifts presented in \cite{Khuskivadze2002}, which depend on the collision energy of the scattering partners. The collision energy is approximated via the semi-classical electronic momentum $k(R)=\sqrt{2/R-1/n_h^2}$, where $n_h$ is the principal quantum number of the closest upper hydrogenic manifold. To properly evaluate the scattering, the electronic wave functions have to be expanded as spherical waves around the position of the ground state atom. Equivalently this can be thought of as a frame transformation onto the scattering center such that a diagonal scattering matrix is obtained. The pseudo-potential is proportional to the scattering length and volume associated with each scattering channel:
\begin{equation}
\hat{V}_{\text{P}}(\v{R},\v{r}) = \sum_{\beta} \frac{(2L+1)}{2} a(S,L,J,k) \frac{\delta(x)}{x^{2(L+1)}} \dyad{\beta}.
\end{equation}
Here, $x=|\v{r}-\v{R}|$ is the relative distance between the Rydberg electron and the ground state atom and $\beta=(S,L,J,M_J)$ labels angular momentum quantum numbers in the frame of the ground state atom, with the projection of total angular momentum $M_J$ of the Rydberg electron and valence electron pair relative to the ground state atom. Note that the above representation includes the change of basis from the unperturbed Rydberg atom to the ground state atom. This is described more elaborately in reference \cite{Eiles2017}. Asymptotically, when the ground state atom is sufficiently far from the Rydberg electron, the PES can be classified by the Rydberg electron's angular momentum relative to the ionic core $j$ and the total spin $F$ of the ground state atom. $\alpha/2R^4$ is the polarization potential induced by the ionic core of the Rydberg atom, with polarizability $\alpha=319.2$ a.u. \\
Lastly, $\hat{H}_{\text{B}}$ describes the impact of the magnetic field, which we assume to be homogeneous. It couples to all angular momenta and spins of the system. For sufficiently weak fields only the paramagnetic or Zeeman terms are important such that
\begin{equation}
\hat{H}_{\text{B}} = \frac{\v{B}}{2} \cdot \left( \hat{\v{j}} + \hat{\v{s}}_1 + 2\hat{\v{s}}_2 \right).
\end{equation}
Since the coupling to the nuclear spin of the ground state atom is suppressed by $1/M$, where $M$ is the nuclear mass, it can be safely neglected. The quadradic Zeeman effect is suppressed for the principle quantum numbers and field strengths considered in this study. The ratio of quadradic and linear effect scales with $n^4B$, where $B$ is given in units of $2.35 \times 10^9$ G \cite{gallagher2005}. We choose the magnetic field to be parallel to the $z$-axis $\v{B} = B \vu{e}_z$. Due to the azimuthal molecular symmetry corresponding to rotations of the internuclear axis around the z-axis, the internuclear axis can without loss of generality be chosen as
\begin{equation}
\v{R} = R (\sin \theta \ \vu{e}_x + \cos \theta \ \vu{e}_z).
\end{equation}
$\hat{H}(\v{r};\v{R})$ is diagonalized for a truncated basis set $\ket{nljm_jm_2m_I}$ ($I$ and $s_2$ are constant and therefore neglected here), where $n$ and $l$ is fixed to a single value (unless indicated otherwise), $j=l\pm1/2$, and all other quantum numbers are covered completely. $\v{R}$ is discretized on a spatial grid thus obtaining $\epsilon_i(\v{R}) = \epsilon_i(R,\theta)$ which is then used in the stationary vibrational Schrödinger equation
\begin{equation}
\left( \v{P}^2/M + \epsilon_i(\v{R}) \right) \Phi_\nu(\v{R}) = E_\nu \Phi_\nu(\v{R}),
\end{equation}
with the vibrational energies $E_\nu$ and eigenfunctions $\Phi_\nu(\v{R})$. This set of partial differential equations is solved via a 2D finite difference method in cylindrical coordinates \cite{kurz_ultralong-range_2014} to obtain the molecular vibrational states and energies crucial for the determination of the binding properties. Beyond this, they are of course of immediate relevance to the analysis of observational spectroscopic data.

\section{Field-free Potential Energy Curves \label{Fieldfree}}

\begin{figure}
\centering
\includegraphics[scale=1]{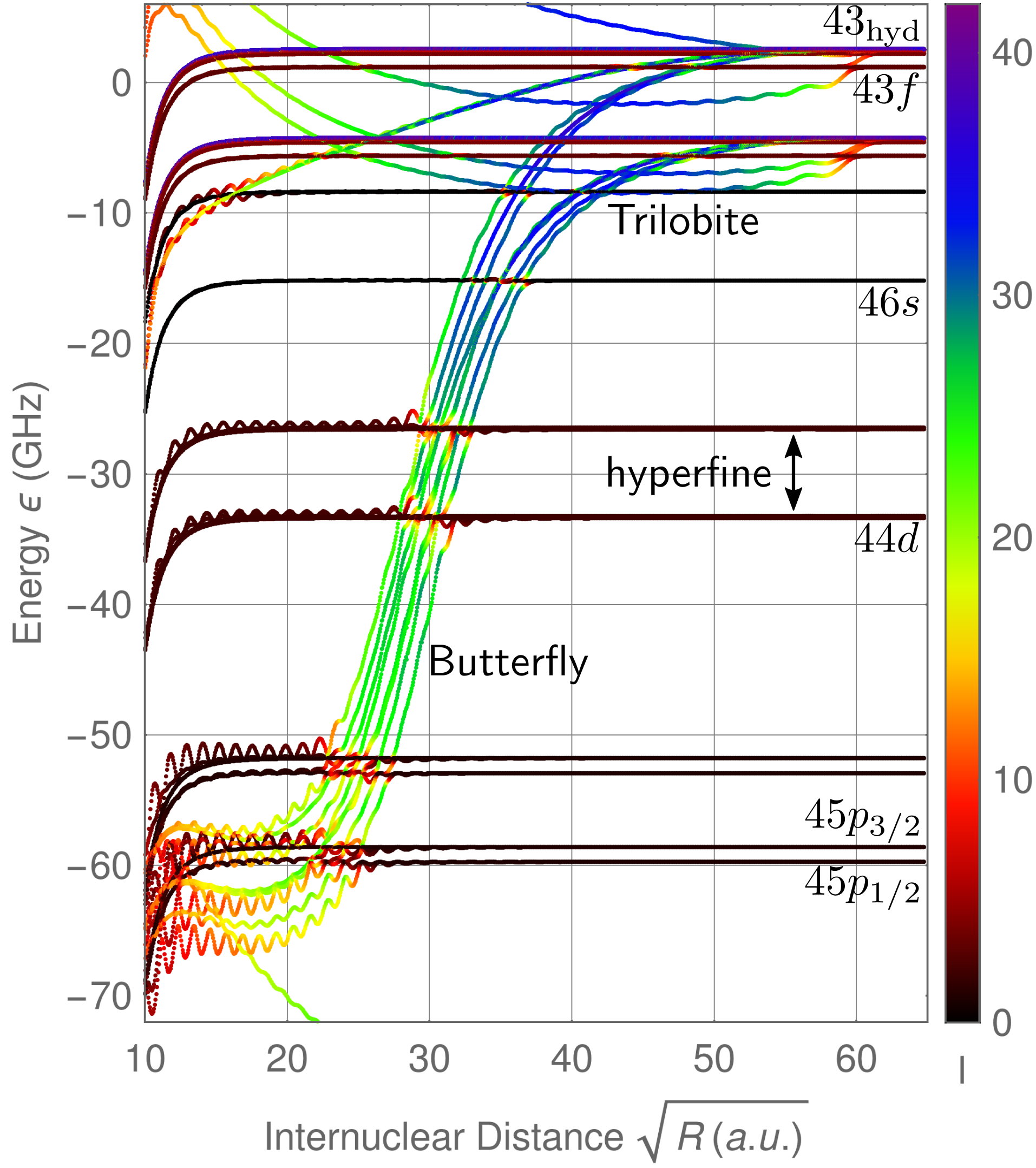}
\caption{Complete manifold of the $n=43$, $\Omega=1/2$ PES with nearby quantum defect split states of low angular momentum in field-free space. The internuclear axis is chosen to coincide with the z-axis. Each PES for a given angular momentum $l$ splits into two hyperfine states $F=2$ and $F=1$. The PES of $p$-states are additionally substantially spin-orbit split. At the top of the figure is the high $l$ hydrogenic manifold from which the trilobite PES emerge. PES cutting through the entire spectrum stem from the p-wave resonant scattering and are referred to as butterfly states. The colorcode indicates the angular momentum $l$-character of the PES. \label{fieldfree}}
\end{figure}

In the field-free case, the PES depend only on the internuclear distance R: $\epsilon_i(\v{R}) \rightarrow \epsilon_i(R)$. Then, without loss of generality, we can choose the internuclear axis to be the z-axis. The projection of the total spin $\Omega = m_j+m_2+m_I$ onto the internuclear axis is an exact conserved quantum number of the molecular system \cite{Eiles2017}. \\
Figure \ref{fieldfree} shows the PES and the Rydberg electron's angular momentum $l$ for principal quantum number $n=43$. Most of the PES come in pairs of total spin $F=2$ and $F=1$ with an energy gap of approximately 7 GHz, due to the hyperfine splitting for rubidium in its ground state. This behavior is exact within our model and most prominent for large distances R, where the two atoms do not interact. Coming from large distances $R$, the PES belonging to the high angular momentum polar states \cite{greene_creation_2000} deviate from the hyperfine split hydrogenic manifold close to zero energy by approximately 4 GHz and can be explained by pure s-wave scattering. These trilobite curves are crossed by $l=3$ $f$-states with a small quantum defect. For lower energies we encounter the PES of the $s$-, $d$-, and $p$-state, respectively. The $p$-state comprises slightly different quantum defects for $j=3/2$ and $j=1/2$ due to spin-orbit effects, introducing an additional visible splitting. For the $d$-state this splitting is smaller and practically not visible on the energy scale of figure \ref{fieldfree}. The curves originating in the hydrogenic manifold and cutting through all other PES are the butterfly curves. Their steepness originates from a resonance of the p-wave scattering, when the phase shift is $\pi/2$ for certain electronic kinetic energies. Here, multiple curves are visible for each hyperfine state, which is due to the spin-orbit interaction of the scattering partners. In the proximity of the Rydberg core $\sqrt{R}\lesssim 15$, the polarization potential becomes dominant. Figure \ref{fieldfree} was derived by diagonalizing $H(\v{r};\v{R})$ for discrete values of the internuclear distance $R$ and using a truncated basis set for the principal quantum numbers $\{n-2,\ldots,n+1\}$, where $n=43$. Since for scattering of the Rydberg electron off the ground state atom only partial waves up to p-wave (i.e.~$L\leq1$) are included, only states with $|m_j|\leq3/2$ will be perturbed. $H(\v{r};\v{R})$ is block diagonal in $\Omega$ and $\Omega=1/2$ is shown in figure \ref{fieldfree}. \\
The spherical symmetry of the PES in the diatomic case can be broken when several ground state atoms are located in the vicinity of the Rydberg atom and polyatomic bound states are possible. For the triatomic case, this is studied in \cite{Fey2016}. In our work, the presence of a magnetic field causes the breaking of the spherical symmetry.

\section{D-state Rydberg Molecules in a Homogeneous Magnetic Field \label{ULRM}} 

\begin{figure}
\centering
\includegraphics[scale=0.95]{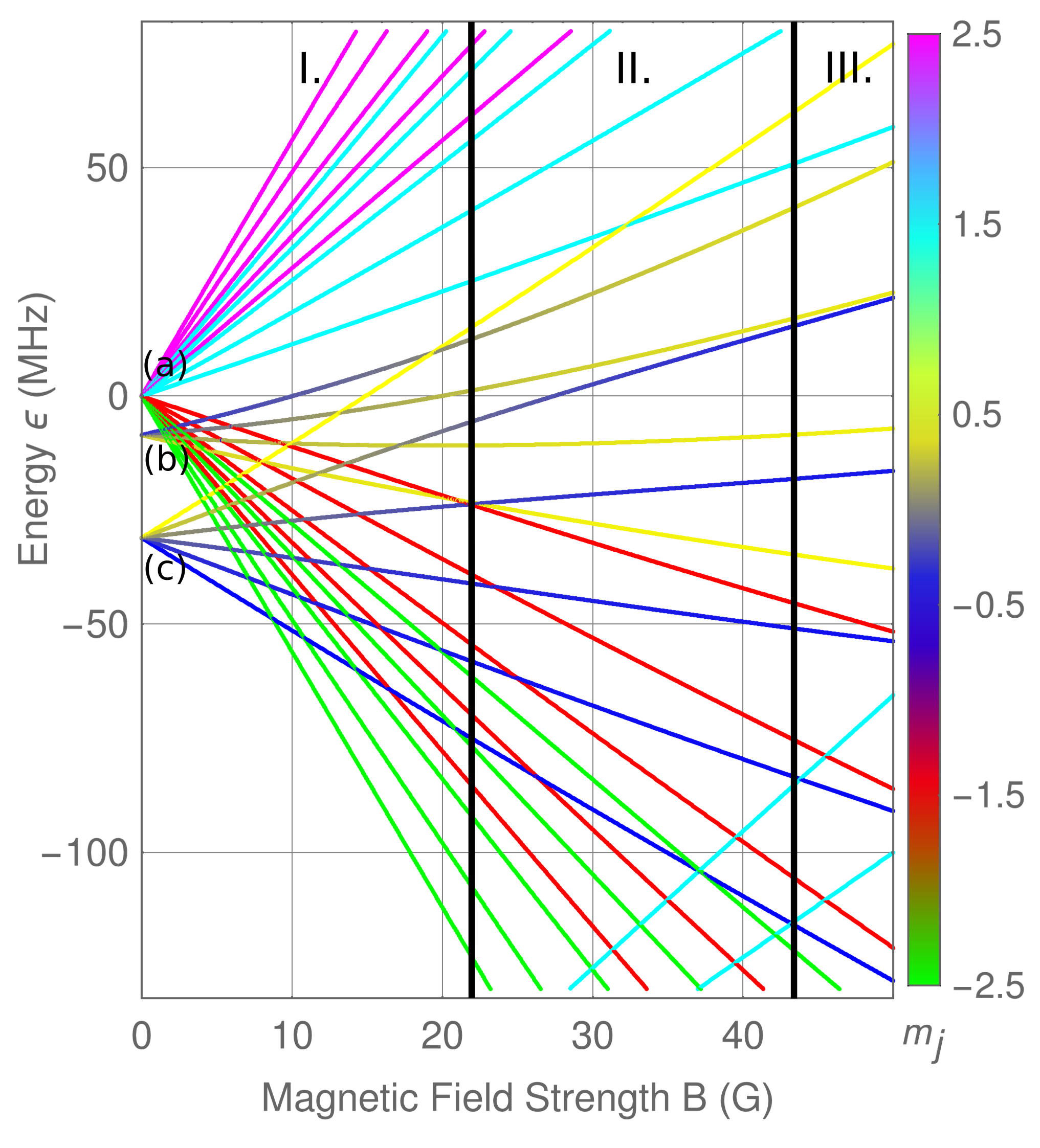}
\caption{Energies of the $38d$-states for $F=2$ and $j=5/2$ with varying magnetic field strength and for a fixed internuclear distance $R=2530$ a.u. This corresponds to the radial position of the outermost potential well. The magnetic field is aligned parallel to the internuclear axis. The colorcode indicates the Rydberg electronic total angular momentum projection $m_j$. Each color of equal $m_j$ appears five-fold due to the five $m_F$ states for $F=2$. $|m_j|>1/2$-states do not couple to the ground state atom positioned on the z-axis, hence the associated curves linearly emerge from zero energy for absent fields (a). $|m_j|\leq1/2$-states are perturbed by the ground state atom (b and c). Three regimes of magnetic field strengths are identifiable with different dominant interactions, namely s-wave scattering (I.), fine structure (II.), and Zeeman splitting (III.). \label{37F}}
\end{figure}

In the following, we focus on $d$-state Rydberg molecules i.e.~with an angular momentum of $l=2$ carried by the Rydberg electron. Therefore, we restrict the basis set also to $l=2$. Diagonalizing an entire hydrogenic manifold is not feasible here, since all total angular momentum projection states have to be taken into account to obtain converged results. However, since the $l<3$-states are detuned from the hydrogenic manifold due to their quantum defects, $l$ is very accurately approximated to be a good quantum number of the molecular system in the vicinity of the outermost potential well. When a magnetic field is present the $(2j+1) \times (2F+1)$-fold degeneracy of each PES in the separate atom limit $R\rightarrow \infty$ is lifted. The asymptotic energy splitting (due to the anomalous Zeeman effect) is given by
\begin{equation}
\Delta \epsilon = B \left[ m_j \left( 1+\frac{(-1)^{l+1/2-j}}{1+2l} \right) +  m_F \frac{(-1)^{2-F}}{2} \right].
\end{equation}
We can distinguish between three regimes of magnetic field strength when compared to other interactions, which are discussed in detail in this section:
\begin{itemize}
\item[I.] The s-wave regime for weak magnetic fields, where the Zeeman splitting is small compared to the effects of the s-wave scattering. In this regime $m_j$ and $m_F$ mixing dominates the shape of the PES.
\item[II.] The fine structure regime for intermediate magnetic fields, where the Zeeman splitting is tuned to exceed the s-wave scattering interaction, but is smaller than the spin-orbit splitting. In this regime $m_j$ and $m_F$ are approximately good quantum numbers.
\item[III.] The Zeeman regime for strong magnetic fields, where the Zeeman splitting is comparable or larger than the fine structure splitting. In this case, the shapes of the PES are strongly tunable since states of different Rydberg electronic angular momenta $j$ mix.
\end{itemize}
This distinction into the three regimes I., II., and III. is shown in figure \ref{37F}. Before we enter into the discussion of the case of the presence of a magnetic field, let us elucidate the zero field case for $d$-state Rydberg molecules: At the position of the outermost potential well, which for the $38d$-state lies at $R=2530$ a.u., three values for the energy for $F=2$ and $j=5/2$-states are possible. (a) Zero energy i.e.~there is no coupling of the Rydberg electron to the ground state atom. Since the internuclear axis coincides with the z-axis, these states carry $|m_j|>1/2$. The absence of the coupling can be understood in terms of the spherical harmonics for $l=2$ and $m_l=\{-l,\ldots,l\}$, which introduce the probability to find the electron at a given angle relative to the z-axis. Only $m_l=0$-states have a non-zero probability on the z-axis. Via the Clebsch-Gordan coefficients, $m_l=0$-states can only contribute to $m_j=\pm1/2$. (b) Slightly lowered energy for states with $|m_j|<1/2$, corresponding to a shallow potential well at $R=2530$. This can be understood in terms of the scattering lengths of the associated spin-dependent s-wave scattering, which is negative for triplet scattering and comparatively small for singlet scattering. The shallow potential well emerges for anti-parallel spin alignment of the scattering partners, namely the Rydberg electron and the valence electron of the ground state atom. Note that $m_1=\pm1/2$ dominantly contributes to $m_j=\pm1/2$. Anti-parallel spin alignment corresponds to a mixed singlet-triplet state, therefore the potential depth can not be maximal. (c) Strongly lowered energy for states with $|m_j|<1/2$, corresponding to a deep potential well and to maximal triplet character i.e.~dominantly parallel alignment of the scattering partners. \\
This s-wave induced energy splitting is complemented by a Zeeman splitting, when a magnetic field is present. For $|m_j|\leq1/2$ mixing of $m_j$ and $m_F$-states emerges. When the internuclear axis is rotated relative to the magnetic field axis (not shown in figure \ref{37F}) also other $m_j$-states start to mix due to the mixing of orbital $m_l$-states to account for the spatial position of the Rydberg electron. \\
For intermediate fields the mixing of states with either different $m_j$ or $m_F$ is suppressed and states of equal $m_j$ appear in sets of consecutive, decreasing $m_F$. For stronger fields the Zeeman splitting exceeds the fine structure and $j$-mixing emerges.

\subsection{The Fine Structure Regime for Intermediate Magnetic Fields \label{intermediate}}

\begin{figure}
\centering
\includegraphics[scale=0.95]{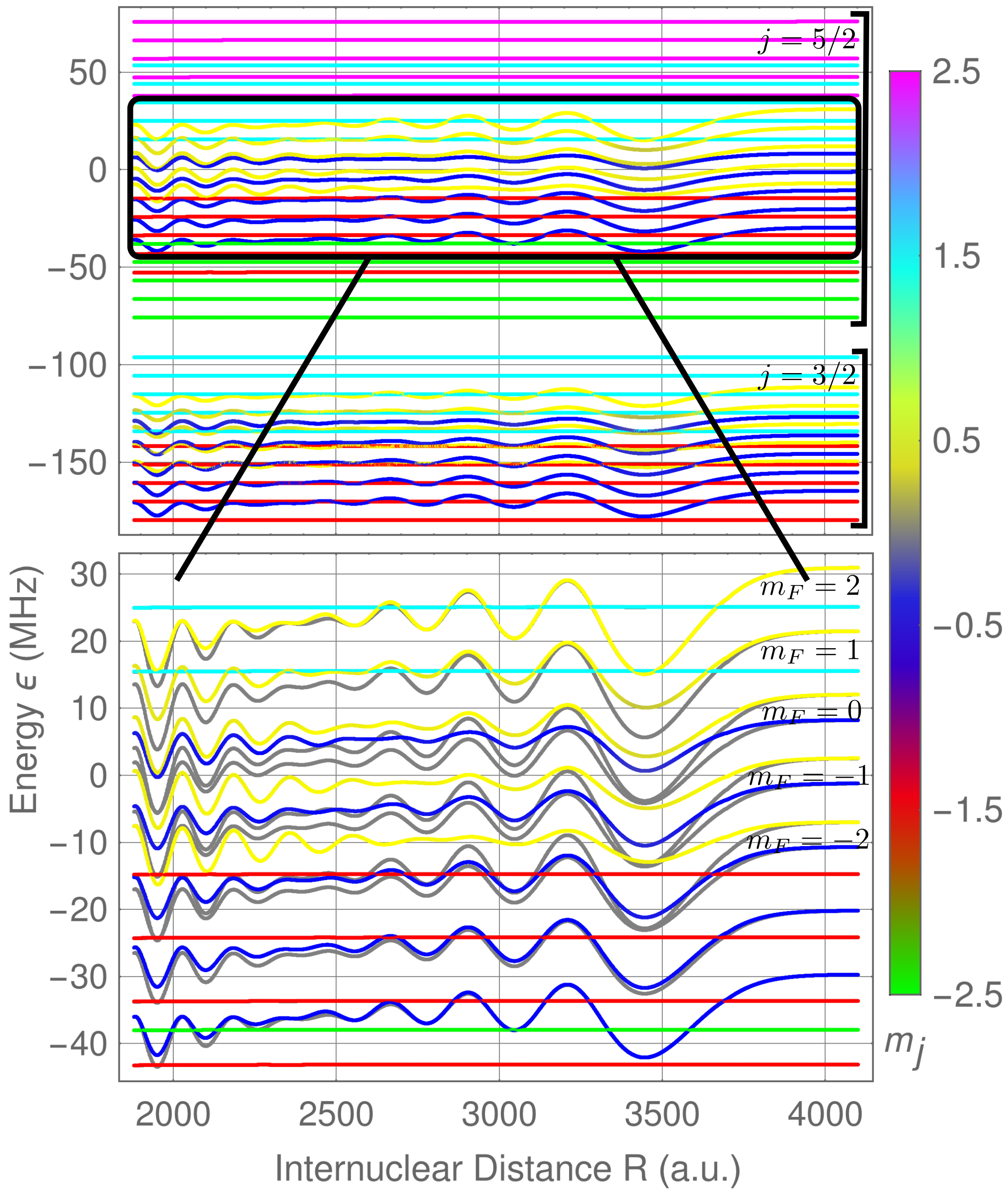}
\caption{Energies of the radial PES for $44d$-states with $F=2$, $\theta=0$, and a magnetic field of $B=13.55$ G. The $j$- and $F$ sublevel degeneracy w.r.t~$m_j$ and $m_F$ is lifted when a magnetic field is present. $m_j$ is indicated by a colorcode. The $j=3/2$ PES are split up into 20 curves, where 5 carry each the same $m_j$ and 4 each the same $m_F$. The energetically higher $j=5/2$ PES is analogously split up $5 \times 6$ times. For $\theta=0$ only states with $|m_j|\leq1/2$ couple to the ground state atom via s-wave interaction and form oscillating PES (bottom). Here, PES from a theory where the electron-neutral atom scattering is assumed to be spin-independent are indicated in grey and deviate significantly. \label{D0}}
\end{figure}

\begin{figure}
\centering
\includegraphics[scale=0.95]{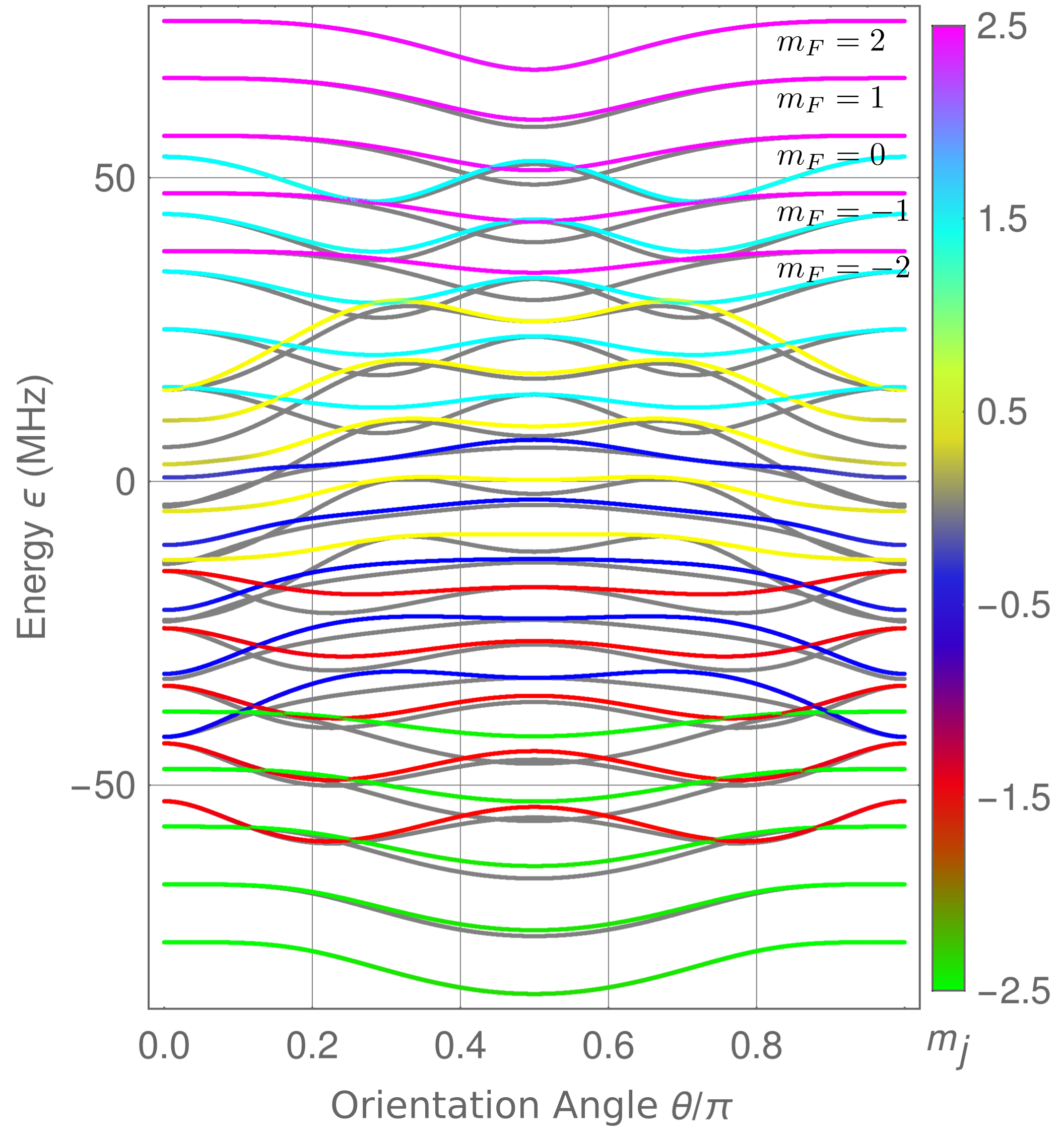}
\caption{Radial cuts of the PES for $44d$-states with $F=2$, $j=5/2$, $R=3452$ a.u., and a magnetic field of $B=13.55$ G with varying angle $\theta$ between the internuclear axis and the magnetic field. Each equally colored PES of constant $m_j$ carries a different nuclear spin projection $-2<m_F<2$ and exhibits potential wells which support molecular states/geometries for different angles $\theta$. For $m_j>0$ ($m_j<0$), the depth of the potential wells decreases for decreasing (increasing) $m_F$. The grey PES are derived upon neglecting the spin dependence in the electron-neutral atom scattering. In this case, the depth of the potential wells is constant for equal $m_j$. \label{AD0}}
\end{figure}

\begin{figure}
\centering
\includegraphics[scale=0.85]{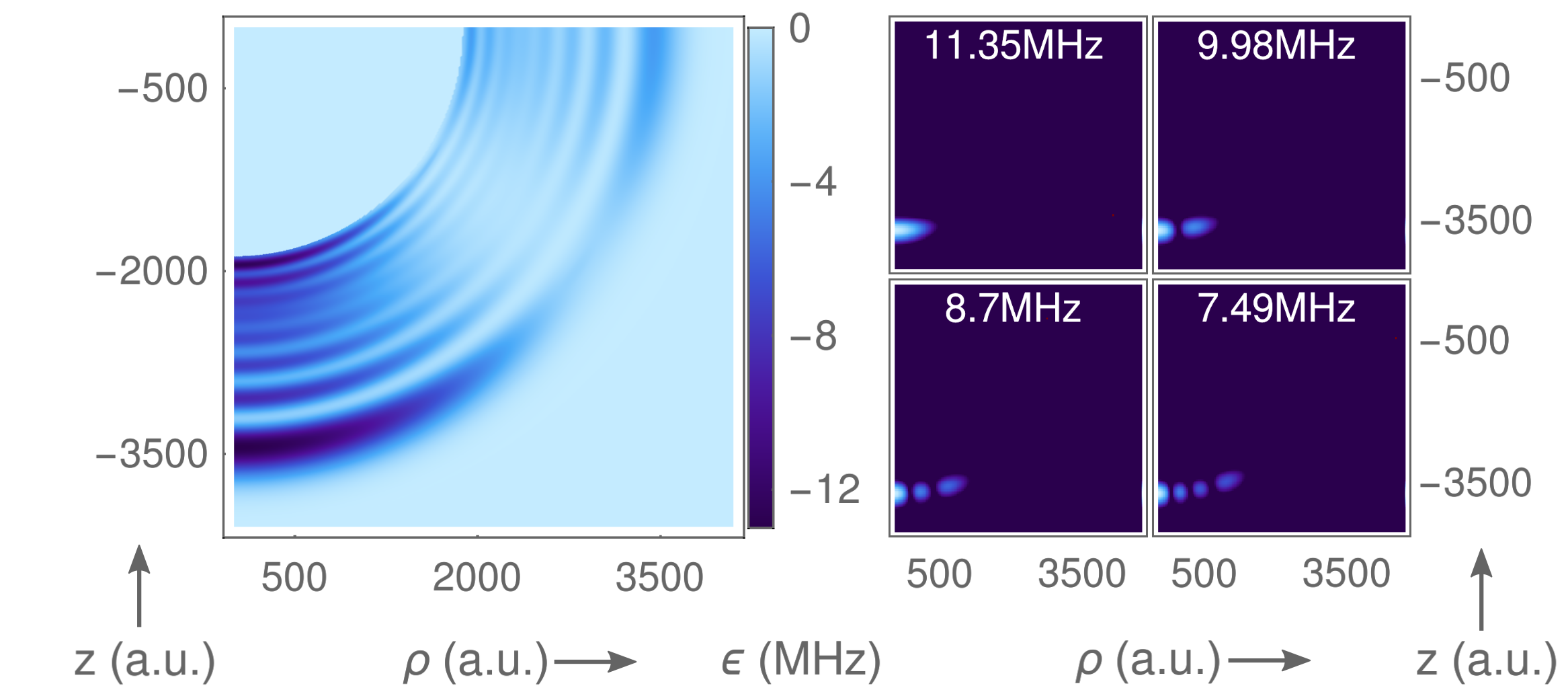}
\caption{2D PES for the $44d$-state with $F=2$, $m_F=2$, $j=5/2$, and $m_j=1/2$ in cylindrical coordinates (left). It supports several molecular vibrational excitations, the four energetically lowest states are shown, along with their binding energies (right, logarithmic color scaling). \label{2DPES}}
\end{figure}

\begin{figure*}
\centering
\includegraphics[scale=1]{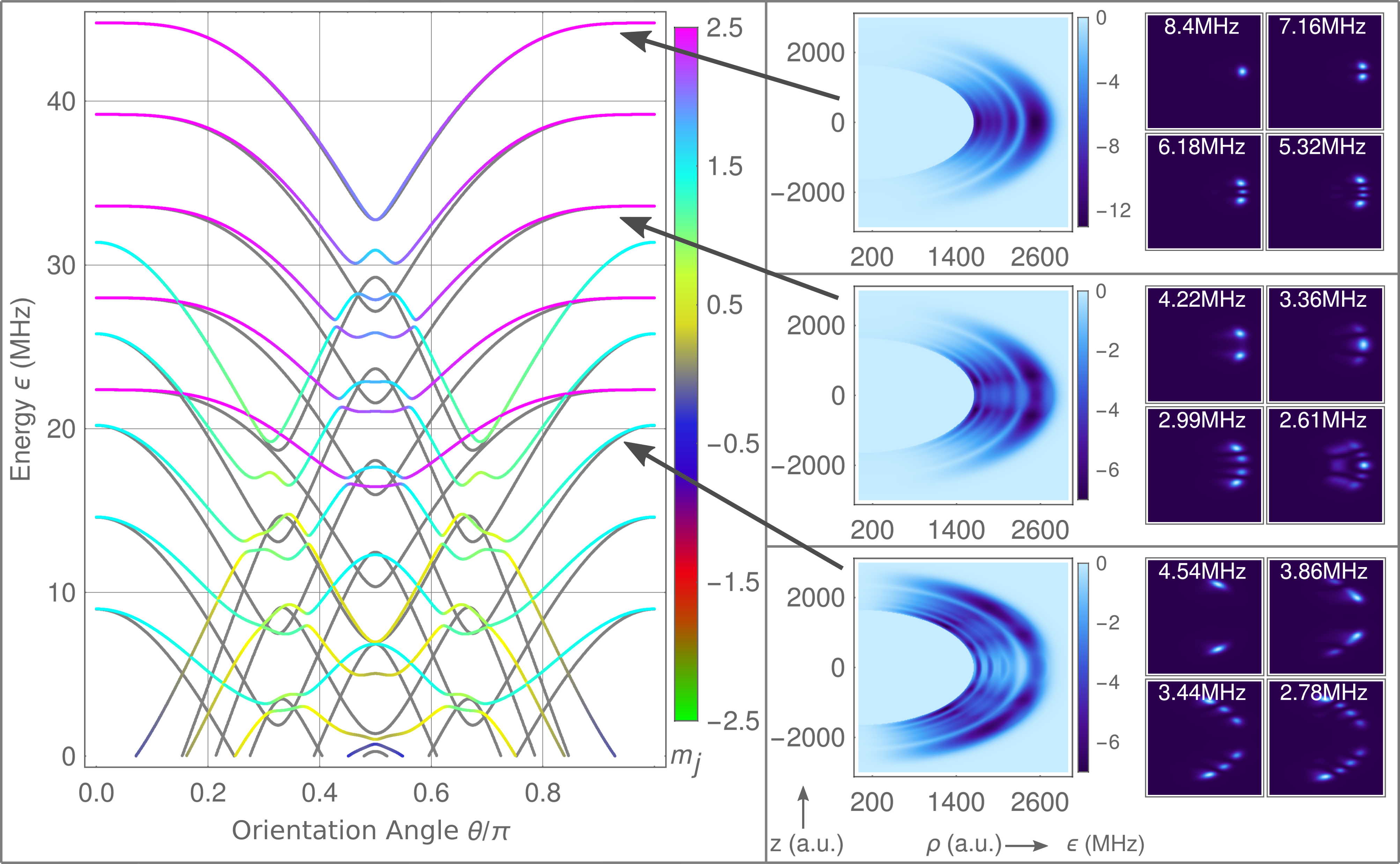}
\caption{Angular dependence of the PES for $38d$-states with $F=2$, $j=5/2$, $R=2530$ a.u., and a magnetic field \mbox{$B=8$ G} - upper part of the spectrum (left). Spin-independent scattering is shown in grey, while a colorcode indicates the $m_j$-state for the model where spin couplings are included. On the right, full 2D PES in cylindrical coordinates are shown for selected states, namely those dominated by their ($m_j=5/2$, $m_F=2$)-character which exhibits a single potential well, ($m_j=5/2$, $m_F=0$)-character with a triple potential well, and ($m_j=3/2$, $m_F=0$)-character (from top to bottom), each accompanied by the normalized densities of the four lowest vibrational molecular states (axes to scale of the 2D-PES), which show a variety of geometric alignments. \label{up37}}
\end{figure*}

We start the discussion of our results with the intermediate regime due to its immediate link to the existing literature. A special case in this regime excluding all spin and spin-orbit interactions was under investigation in \cite{krupp_alignment_2014}. There, only triplet scattering ${}^3S_1$ and ${}^3P_{0}$ was assumed for every channel. In this section, we focus on $n=44$ to match the experimental study  \cite{krupp_alignment_2014}. Furthermore, we investigate $n=38$ for the other regimes below, since a lower principle quantum number leads to deeper potential energy wells and therefore, allows to achieve a better localization of vibrational states in different angular configurations. Figure \ref{D0} shows the PES of the $F=2$ and $44d$-state for a magnetic field of $B=13.55$ G and fixed orientation angle $\theta=0$. The PES are characterized by a colorcode according to the total angular momentum projection of the Rydberg electron $m_j$. In principle, due to the magnetic field and electronic scattering, states of different spin can mix. However, in the intermediate regime the energetic separation of different spin-states exceeds the effects of the scattering. Therefore, $m_j$ and $m_F$ are approximately good quantum numbers in this regime. Particularly, $\Omega$ is approximately conserved also for non-zero $\theta$. For all other regimes we consider in the work, $\Omega$ is conserved only for $\theta=0$. For the given magnetic field in figure \ref{D0}, the Zeeman splitting is visible for $j=3/2$ and $j=5/2$ which are energetically clearly separated. Each color appears 5-fold due to the 5 possible $m_F$ states. As described in earlier parts of section \ref{ULRM}, for the given orientation angle $\theta=0$, only PES of $m_j=\pm 1/2$-states interact with the ground state atom. In contrast to the field-free case, where the PES of $m_j=\pm 1/2$-states comprise either a deep or a shallow outermost potential well, for intermediate field strength this distinction obliterates. The outermost well varies in depth for different $m_F$. For $m_j=1/2$ it deepens for increasing $m_F$ and vice versa for $m_j=-1/2$, as visible in the lower part of figure \ref{D0}. For intermediate fields, the energies of the $m_j$ and $m_F$-states are not degenerate anymore, and the variation of the potential depth can be directly associated with the spin alignment of the Rydberg and valence electron of the ground state atom: In $m_j=1/2$ ($m_j=-1/2$)-states, the dominant electronic spin contribution is $m_1=1/2$ ($m_1=-1/2$). Simultaneously, the $m_F=2$-state is a pure electronic $m_2=1/2$-state, therefore resulting in maximal triplet (mixed) character. The contribution of $m_2=1/2$ gradually decreases for decreasing $m_F$, such that $m_F=-2$ is a pure electronic $m_2=-1/2$ state. The shallower wells mirror the change from a triplet-dominant character to a mixed-state-dominant character. In a model where the electron-neutral ground state atom scattering is assumed to be spin-independent the outermost potential wells possess a constant depth. In figure \ref{D0} the associated PES to the latter model are shown in grey. \\
Under variation of $\theta$, more features of the PES become apparent and are presented in figure \ref{AD0}, where $\theta$ is varied and the internuclear distance is fixed to the position of the outermost potential well $R=3452$ a.u. The potential wells emerge approximately for three different angular positions. These are $\theta=0$ and $\theta=\pi$ for $m_j=\pm 1/2$-states, $\theta=\pi/4$ and $\theta=3\pi/4$ for $m_j=\pm 3/2$-states, and $\theta=\pi/2$ for $m_j=\pm 5/2$-states. This can be understood in terms of the above mentioned spherical harmonics for $l=2$ and $m_l=\{-l,\ldots,l\}$, which have angular positions of maximal electronic density at the orientation angles $\theta=0$ for $m_l=0$, $\theta=\pi/4$ for $m_l=\pm1$, and  $\theta=\pi/2$ for $m_l=\pm2$ and contribute to $m_j$-states accordingly. Equally to the feature seen in figure \ref{D0}, the depth of the potential wells decreases for decreasing (increasing) $m_F$ when $m_j>0$ ($m_j<0$), whereas for the spin-independent model the depth is constant. \\
These deviations are particularly interesting, as the $m_j=5/2$, $m_F=2$-state (upper most pink PES in figure \ref{AD0}) and the $m_j=1/2$, $m_F=2$-state (upper most yellow PES in figure \ref{AD0}) were investigated in reference \cite{krupp_alignment_2014}. While the PES for the $m_j=5/2$, $m_F=2$-state are practically identical in the spin-independent model used in \cite{krupp_alignment_2014} and the spin-dependent model used in this work, the PES for the $m_j=1/2$, $m_F=2$-state is narrower around the molecular ground state at $\theta=0$. A full two-dimensional image of the latter particular PES is depicted in figure \ref{2DPES}, along with the four energetically lowest vibrational states and binding energies. The narrower potential well leads to an increased energy spacing of bound vibrational states. In fact, discrepancies between the energetic separation of experimental and theoretical states have been observed in reference \cite{krupp_alignment_2014}. Our results may explain these discrepancies and attribute them to the included spin couplings. We were able to reduce the relative energetic deviation from the spectrum from around $10\%$ in the spin-independent model to less then $5\%$ when spin couplings are included, and the average relative energy spacing of the vibrational states from $4\%$ to $2\%$.

\subsection{The s-wave Regime for Weak Magnetic Fields \label{weak}}

\begin{figure*}
\centering
\includegraphics[scale=1]{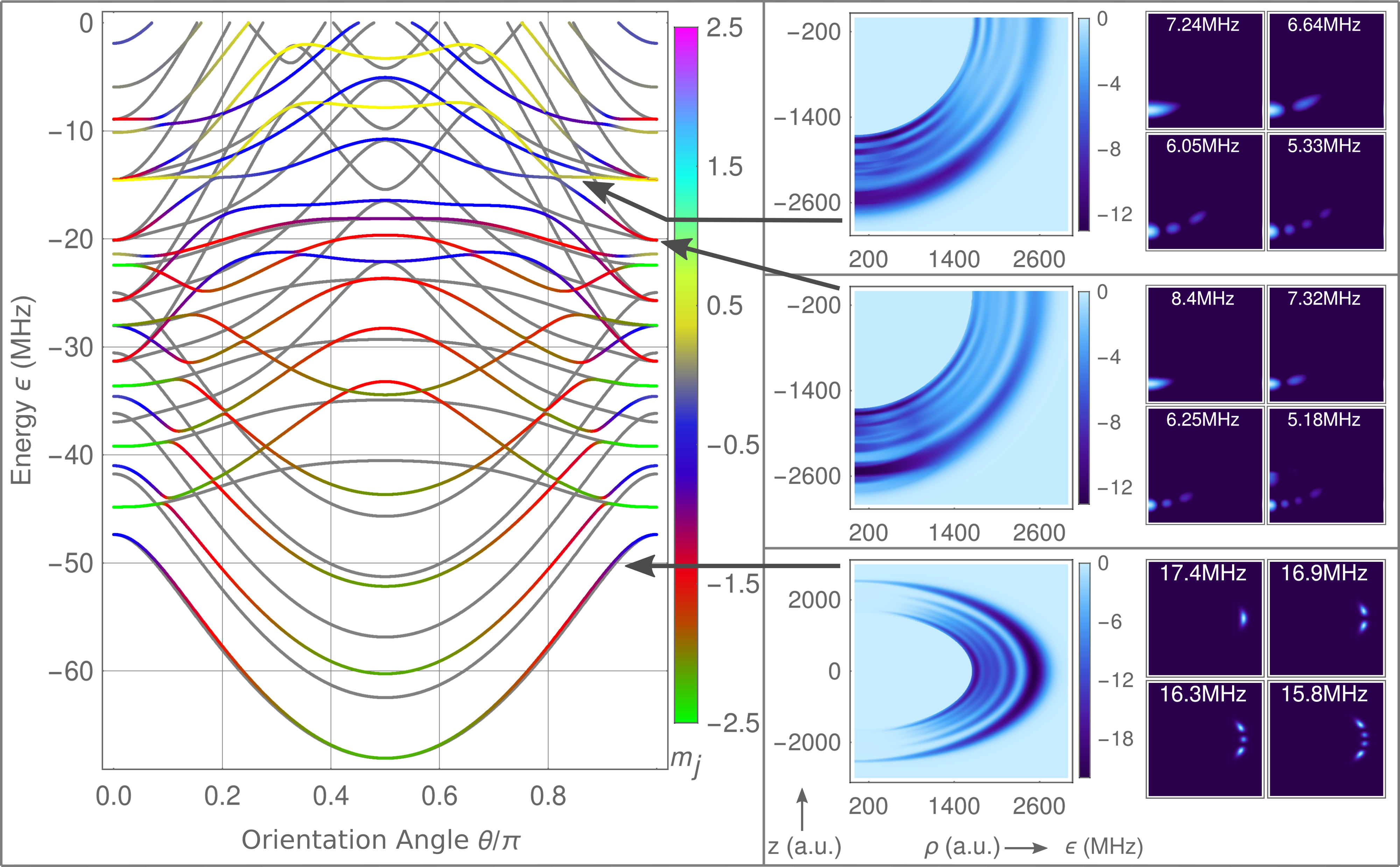}
\caption{Angular dependence of the PES for $38d$-states with $F=2$, $j=5/2$, $R=2530$ a.u., and a magnetic field $B=8$ G - lower part of the spectrum (left). The focus here is on states dominated by ($m_j=1/2$, $m_F=2$), ($m_j=-1/2$, $m_F=0$) and ($m_j=-5/2$, $m_F=-2$)-character. Exemplarily, their full PES in cylindrical coordinates are shown (right). The latter displays a multitude of excitations (axes to scale of the 2D-PES) purely in angular direction, while in the former case only faint angular excitations are present (logarithmic color scaling in the two upper plots). \label{low37}}
\end{figure*}

When the magnetic Zeeman splitting of states of different total angular momentum projection $m_j$ and total nuclear spin projection $m_F$ is weak compared to the effect of the interaction with the ground state atom, the PES may strongly influence each other allowing for avoided crossings which lead to molecular vibrational eigenstates for new geometries and spin states. The emerging complex patterns are depicted in figures \ref{up37} and \ref{low37}. As an example, we focus on a Rydberg atom in the $38d$-state in a magnetic field $B=8$ G. Although an overall pattern as in the intermediate regime is still visible, namely that potential wells become shallower as $m_F$ decreases (increases) for $m_j>0$ ($m_j<0$), the PES acquire novel features due to their interaction with energetically closeby PES. States with $m_j=5/2$, which comprise a stable molecular geometry when the internuclear axis is perpendicular to the magnetic field in the intermediate regime, gain a richer structure with double ($m_F=1$) and triple ($m_F=0$) wells around $\theta=\pi/2$. This is also true for states of $m_j=3/2$ with local minima at $\theta=\pi/4$ and $\theta=3\pi/4$. As seen in figure \ref{up37}, the richer potential well structure leads to novel vibrational states with varying equilibrium positions and degree of localization. \\
Additionally, for $m_j=\pm1/2$-states, the typical structure of groups of PES with the same $m_j$ and decreasing $m_F$ breaks down in the weak field regime. This succession is disrupted by the s-wave interaction and can be explained in terms of the strong influence of the singlet and triplet $S$-state which governs the scattering. This is also visible in figure \ref{37F} I. However, the general shape and depth of the potential wells is not influenced much by this effect. \\
Opposite to this, the energetically lowest PES of $j=5/2$ shown in figure \ref{low37} exhibits a surprisingly deep potential well at angular position $\theta=\pi/2$. Pictorially speaking, this PES inherits the features of the $m_j$ state, that minimizes the energy at a given orientation angle: $m_j=-1/2$ for $\theta \approx 0$, $m_j=-3/2$ for $\theta \approx \pi/4$, and $m_j=-5/2$ for $\theta \approx \pi/2$. This effect makes the arising potential well twice as deep as it would be without $m_j$ mixing. Thus, molecular states are relatively deeply bound with vibrational excitations purely in angular direction.

\subsection{The Zeeman Regime for Strong Magnetic Field \label{strong}}

\begin{figure*}
\centering
\includegraphics[scale=0.90]{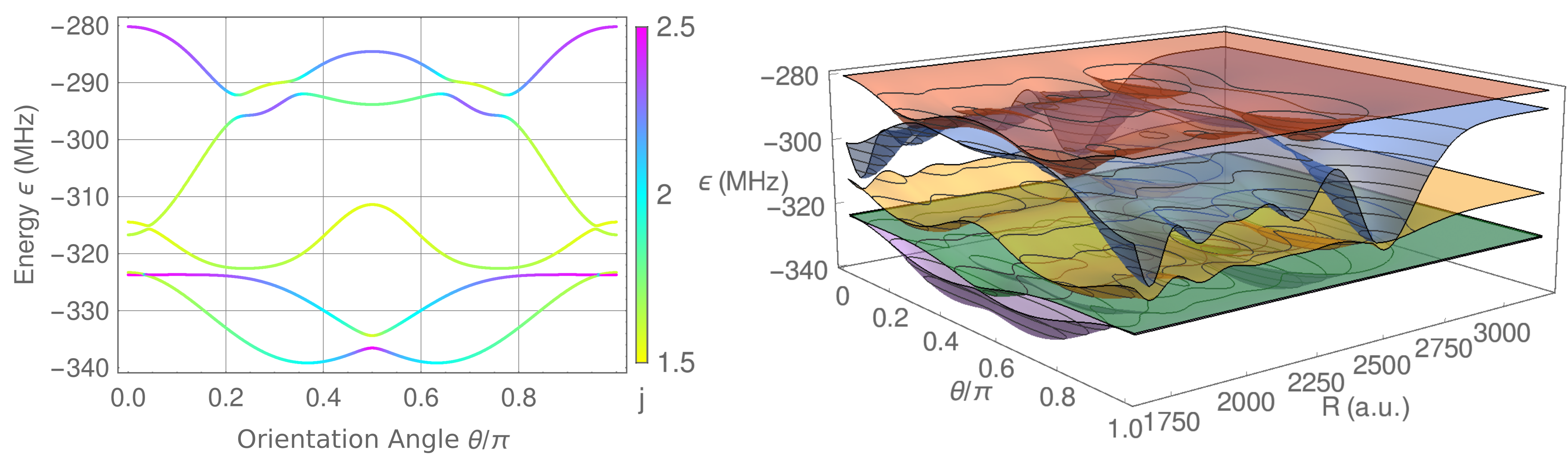}
\caption{Left: Angular dependence of the PES for $38d$-states in a magnetic field $B=93$ G at $R=2530$ a.u. Due to the large field only a small subset of PES are shown. The color indicates the $j$-character of the PES, which strongly mixes in this regime of field strength. The spin-orbit separation is overcome and energetically close PES influence each other, resulting in molecular states at various orientation angles $\theta$ (lowest and uppermost curve) and with tunable depth at the most common well positions $\theta=0$ (second uppermost curve) or $\theta=\pi/2$ (not visible here). Right: Full 2D surfaces of the same PES. \label{37B93}}
\end{figure*}

For a Zeeman splitting that exceeds the spin-orbit splitting, even more complex PES emerge due to mixing of states with different total angular momentum $j$ (compare figure \ref{37B93}). Here, the magnetic field can be used to energetically connect PES of desired angular momentum projections. In the regime of intermediate magnetic fields (II.), with its typical fan-like structure for varying field (see figure \ref{37F}), the energetic distance between PES within the same fan can never vanish, since the s-wave interaction has too little influence on the shape of the PES. Thus, for increasing fields, the relative impact that PES in the same fan have on each other, can only decrease. In contrast to this, in the strong field regime, the magnetic field can be used to combine states (and PES) from different $j$ almost arbitrarily, since they span another fan. This results in avoided crossings which allow PES with exotic shapes. Since for different $m_j$, the curves have potential wells at different orientation angles $\theta$, the emergent $m_j$-mixing makes the angular position of the resulting potential wells tunable, so that molecules can form in desired geometric configuration relative to the magnetic field. Additionally, by combining curves with different $j$, but equal $m_j=0$, the depth of the potential well at $\theta=0$ can be increased. This is also possible when combining $j=5/2$, $m_j=\pm5/2$-states with $j=3/2$, $m_j=\pm3/2$-states, whose associated PES both comprise a potential well at $\theta=\pi/2$. In figure \ref{37B93}, all of the above mentioned effects are visible as well as a global view on the two-dimensional PES.

\section{Conclusion \label{conc}}

We have explored the impact of a homogeneous magnetic field on ultralong-range Rydberg molecules exemplarily for $d$-states. In contrast to previous works, all relevant spin couplings up to p-wave interactions have been taken into account, including the spin-orbit coupling of the interaction between the ground state atom and the Rydberg electron. The magnetic field couples (otherwise isolated) potential energy surfaces and a rich landscape of surfaces emerges. We were able to identify three regimes of magnetic field strength relative to the strength of the s-wave interaction and the spin-orbit splitting of the Rydberg atom. These three regimes were analyzed regarding equilibrium positions, binding energies, and molecular orientation relative to the magnetic field, quantities that are in principal accessible in experiments. Additionally to the known orientations of $d$-state molecules, which are $\theta=0,\ \pi/4$, and $\pi/2$, the magnetic field in combination with spin couplings gives rise to novel equilibrium configurations at intermediate angles. This is accomplished by controlling the mixing of the angular momentum and spin quantum numbers $j$, $m_j$, and $m_F$. The arising PES of these mixed states comprise complex structures and are strongly tunable. Hence, via the controllable spin content of a molecular state its molecular geometry can be tailored and both properties are strongly intertwined. Parallel as well as perpendicularly aligned ULRM with $l=2$ and $m_F=2$ have already been observed experimentally \cite{krupp_alignment_2014}. Their orientations were determined, however, only indirectly by identifying peaks in the experimental spectrum with vibrational energies of oriented molecular states from theoretical computations. By the same means, new orientation angles proposed in this study can be confirmed experimentally. An alternative option for a more direct experimental measurement would be to drive microwave transitions between vibrational molecular states. The corresponding Franck-Condon factors of these transitions contain information of the spatial arrangement of the vibrational states and therefore depend directly on the relative molecular orientations. \\
The strong coupling of the PES suggests interesting dynamical processes. In a polyatomic setting the magnetic field could make the molecular geometry tunable, leading to exotic molecular shapes. This study can be extended to Rydberg molecules consisting of cesium and strontium as well as to $p$-state molecules in magnetic fields. Here, p-wave interaction plays a significant role and the effect of ${}^3P_J$ splitting is more prominent than for $d$-state molecules. 

\begin{acknowledgements}
The authors would like to thank Matthew T. Eiles for fruitful discussions. F.H. and P.S. acknowledge support from the Deutsche Forschungsgemeinschaft (DFG) within the Schwerpunktprogramm 1929 Giant Interactions in Rydberg Systems (GiRyd). C.F. gratefully acknowledges a scholarship by the Studienstiftung des deutschen Volkes.
\end{acknowledgements}

\bibliographystyle{apsrev4-1}
\bibliography{biblio.bib}
\end{document}